\documentclass[reprint,
showpacs,
 amsmath,amssymb,
prl,
]{revtex4-1}

\usepackage{graphicx}% Include figure files
\usepackage{dcolumn}% align table columns on decimal point
\usepackage{longtable}% align table columns on decimal point
\usepackage{lipsum}% align table columns on decimal point
\usepackage{color}% bold math
\begin{document}

%\preprint{APS/123-QED}

\title{ Quantization of Conductance Minimum and Index Theorem}
%\altaffiliation[Also at ]{}%Lines break automatically or can be forced with \\
\author{Satoshi Ikegaya$^{1}$}
\author{Shu-Ichiro Suzuki$^{1}$}
\author{Yukio Tanaka$^{2,3}$}
\author{Yasuhiro Asano$^{1,3,4}$}
%\email{\empty}
\affiliation{$^{1}$Department of Applied Physics,
Hokkaido University, Sapporo 060-8628, Japan\\
$^{2}$Department of Applied Physics, Nagoya University, Nagoya 462-8602, Japan\\
$^{3}$Moscow Institute of Physics and Technology, 141700 Dolgoprudny, Russia\\
$^{4}$Center of Topological Science and Technology,
Hokkaido University, Sapporo 060-8628, Japan
}%

\date{\today}% It is always \today, today,

\begin{abstract}
We discuss the minimum value of the zero-bias differential conductance $G_{\textrm{min}}$ 
in a junction consisting of a normal metal and a nodal superconductor preserving time-reversal 
symmetry. 
Using the quasiclassical Green function method, we show that $G_{\textrm{min}}$ is quantized at 
$ (4e^2/h) N_{\mathrm{ZES}}$ in the limit of strong impurity scatterings in the normal metal.
The integer $N_{\mathrm{ZES}}$ represents the number of perfect transmission 
channels through the junction. An analysis of the chiral symmetry of the
Hamiltonian indicates that $N_{\mathrm{ZES}}$ corresponds to the Atiyah-Singer index 
in mathematics. 
\end{abstract}


\maketitle

%\tableofcontents
%*********************************************************************************
%\section{Introduction}
%*********************************************************************************

The quantization of an observable value in physics is closely related some of the time to 
an invariant in mathematics.
A good example may be the quantized Hall conductivity in condensed matter physics.
Although the quantization of the Hall conductivity itself occurs for physical reasons,
the quantized value is given by the Chern invariant in a two-dimensional manifold~\cite{tknn}.
Another example is the number of gapless states at the surface of a topologically nontrivial material 
 characterized by a topological invariant $Z$. 
The integer $Z$ depends on the spatial dimensionality and the symmetry class of the
Hamiltonian~\cite{altland,schnyder}. The conductance in a junction consisting of such a 
topologically nontrivial superconductor is quantized at $(2 e^2/h)Z$ with $Z=1$ for a 
one-dimensional class D superconductor~\cite{lutchyn,oreg,mourik,deng,das}. 
A similar phenomenon has been discussed as regards superconductors in class BDI~\cite{tewari,niu,diez,dahlhaus} 
with $Z$ being an integer number. 

The Atiyah-Singer theorem relates an topological invariant to an invariant defined 
in terms of solutions of a differential equation.  
The index theorem provides the mathematical background to the quantum anomaly in particle physics.
In condensed matter physics, the index theorem describes the number of gapless modes 
at a boundary between two chiral superfluids~\cite{volovik}.
Generally speaking, a close relationship between a quantized physical value and a mathematical invariant 
implies the universality of the corresponding phenomenon.
In this paper, we show the relationship between the minimum value of the conductance
in a superconducting junction and the Atiyah-Singer index.

We discuss the zero-bias differential conductance $G_{\mathrm{NS}}$ in a normal-metal/
superconductor (NS) junction in two dimensions, where the normal metal contains a number of 
random impurities and its normal resistance is $R_{\mathrm{N}}$. 
The superconductor is characterized by 
unconventional time-reversal pairing symmetries such as $p_x$-, $d_{xy}$-, and $f$-wave symmetry.
The analytical expression of the conductance is obtained by solving the quasiclassical 
Usadel equation~\cite{eilenberger,larkin,usadel} in a normal metal with an appropriate boundary 
condition at an NS interface~\cite{yt03,yt04,yt05bc}. 
We find that $G_{\mathrm{NS}}$ decreases to the quantized value 
of $(4e^2/h) |N_{\mathrm{ZES}}|$ with increasing in $R_{\mathrm{N}}$.
The integer $|N_{\mathrm{ZES}}|$ is the number of perfect transmission channels 
in a dirty NS junction.
The analysis in terms of the chiral symmetry of the Hamiltonian~\cite{ms1dwn,si15} enables us to understand 
the relationship between $N_{\mathrm{ZES}}$ and the index in the Atiyah-Singer theorem.

%========================================
% \section{Hamiltonian}
%=========================================

\begin{figure}[tbh]
\begin{center}
\includegraphics[width=8.5cm]{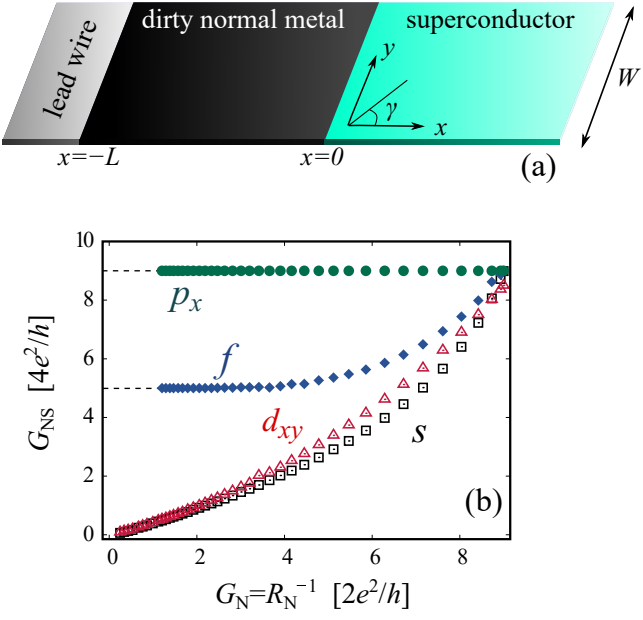}
\end{center}
\caption{(color online). 
(a) Schematic picture of a normal-metal/superconductor junction.
(b) Numerical results on two-dimensional tight-binding model.
The index $|N_{\mathrm{ZES}}|$ is 9 and 5, for $p_x$- and $f$-wave symmetry, respectively.
}
\label{fig1}
\end{figure}

Let us consider a normal-metal/superconductor (NS) junction described by a
$2\times 2$ Bogoliubov-de Gennes (BdG) Hamiltonian,
\begin{align}
\hat{H}_{\mathrm{BdG}} =& \left[ \begin{array}{cc}
\xi_{\boldsymbol{r}} + V(\boldsymbol{r}) & \Delta(\boldsymbol{r}) \\
\Delta(\boldsymbol{r}) & -\xi_{\boldsymbol{r}} - V(\boldsymbol{r})
\end{array} \right], \label{bdg}\\
\xi_{\boldsymbol{r}} =& -\frac{\hbar^2 \nabla^2}{2m} - \mu_F, \\
V(\boldsymbol{r})=&V_{\mathrm{imp}}(\boldsymbol{r})\Theta(-x)\Theta(x+L)
+v_0 \delta(x),\\
\Delta(\boldsymbol{r})=&\left\{ 
\begin{array}{cc}
\Delta, & s\\
-2\Delta\, \partial_x \partial_y/k_F^2, & d_{xy}\\
-i\Delta\, \partial_x / k_F & p_x\\
-i\Delta\, \partial_x( k_F^2 + 2\partial_y^2)  / k^3_F, & f,
\end{array}\right. \label{pair_symmetry}
\end{align}
where $m$ is the mass of an electron, $\mu_F$ is the chemical potential,
$k_F$ is the Fermi wave number and $\Theta(x)$ is the step function. 
We introduce 
the random impurity potential $V_{\mathrm{imp}}$ in the normal metal $(-L<x<0)$ 
as shown in Fig.~\ref{fig1}(a) and consider the barrier potential $v_0$ at the 
NS interface. In the $y$ direction, we apply a periodic boundary condition 
with $W$ denoting junction width.

%

%***************************************
%\section{Usadel equation}
%***************************************

The quasiclassical Usadel equation in the normal metal is represented 
by using $\theta$-parameterization,
\begin{align}
\hbar D \frac{\partial^2 \theta(x,\epsilon)}{\partial x^2} + 2i\, \epsilon \, \sin \theta(x,\epsilon) =0, \label{usadel}
\end{align}
where $D$ is the diffusion constant and $\epsilon$ is the energy of a quasiparticle measured from the Fermi level 
(zero energy).
The normal and anomalous Green functions with retarded causality are given by 
$g=\cos\theta$ and $f=\sin\theta$, respectively. 
The Usadel equation is supplemented by two boundary conditions~\cite{yt03,yt04,yt05bc},
\begin{align}
\theta(x=-L,\epsilon)=0,\quad \frac{L}{G_Q\, R_{\mathrm{N}}} \left. \left(\frac{\partial \theta}{\partial x}\right)\right|_{x=0} = 2 I_F,
\label{bc}
\end{align}
with
\begin{align}
&I_F= {\sum_{k_y}} \frac{ |t_b|^2( f_s \cos\theta_0 - g_s \sin\theta_0)}{\Xi},\label{if_def}\\
&t_b = \cos\gamma/(\cos\gamma + iz_0),\; z_0=mv_0/(\hbar^2k_F) \\
&\Xi= (2-|t_b|^2)z_s + |t_b|^2( g_s \cos\theta_0 + f_s \sin\theta_0),\\
 &\theta_0=\theta(x=0,\epsilon),\quad G_Q=2e^2/h,
\end{align}
where $\gamma$ is the angle measured from the $x$ axis as shown in Fig.~\ref{fig1}(a), and the wavenumber 
is given by $k_x=k_F\cos\gamma$ and $k_y=k_F\sin\gamma$.
The pair potentials in Eq.~(\ref{pair_symmetry}) are represented as
$\Delta\sin(2\gamma)$, $\Delta\cos\gamma$, and $\Delta\cos\gamma(1-2\sin^2\gamma)$ for 
$d_{xy}$-, $p_x$- and $f$-wave symmetries, respectively.
The transport channel is characterized by the wave number in the direction parallel 
to the interface $k_y$. Thus ${\sum_{k_y}}$ is the summation over all the propagating channels
and is evaluated as
\begin{align}
{\sum_{k_y}} \to \frac{W}{2\pi} \int_{-k_F}^{k_F} dk_y = \frac{Wk_F}{2\pi} \int_{-\pi/2}^{\pi/2}d\gamma\, \cos\gamma.
\label{kysum}
\end{align}   
The number of propagating channels is calculated as $N_c=[Wk_F/\pi]_{\mathrm{G}}$, where $[\cdots]_{\mathrm{G}}$ 
is the Gauss symbol giving the integer part of the argument.
We have defined as
\begin{align}
\Delta_+=&\Delta(\gamma), \quad \Delta_-=\Delta(\pi-\gamma),\\
g_\pm=& \frac{\epsilon}{\sqrt{\epsilon^2-\Delta_\pm^2}}, \quad
f_\pm= \frac{i\Delta_\pm}{\sqrt{\epsilon^2-\Delta_\pm^2}},\\
g_s=&g_+ + g_-,\quad z_s = 1+g_+g_- +f_+ f_-,\\
f_s=&\left\{ \begin{array}{cc} f_+ + f_- & \textrm{singlet} \\ i(f_+g_- - f_- g_+) & \textrm{triplet},\end{array}\right.\\ 
\bar{f}_s=&\left\{ \begin{array}{cc} i(f_+g_- - f_- g_+) & \textrm{singlet} \\ f_+ + f_- & \textrm{triplet}.\end{array}\right. 
\end{align}
The normal resistance of a potential barrier at $x=0$ is defined as
$R_{\mathrm{B}}=(G_Q N_c T_B)^{-1}$ by using the transmission probability of the barrier 
$T_B=\int_0^{\pi/2} \cos\gamma |t_b|^2$.
The normal resistance of a normal metal is $R_{\mathrm{N}}$, which is 
the inverse of the normal conductance $G_{\mathrm{N}}$. 
The total resistance of an NS junction at a bias voltage $V_{\mathrm{bias}}$ is 
represented by a modified Ohm's law~\cite{yt03},
\begin{align}
&R_{\textrm{NS}}= \frac{1}{G_Q I_B} + \frac{ R_{\mathrm{N}} }{L}
\int_{-L}^0 
\frac{dx}{\cosh^2\left(\textrm{Im}(\theta(x,\epsilon) \right)},\label{ohm}
\end{align}
by putting $\epsilon \to eV_{\mathrm{bias}}$ on the right-hand side of Eq.~(\ref{ohm}).
The first term represents the resistance due to the potential barrier with~\cite{yt05bc}
\begin{align}
&I_B ={\sum_{k_y}} \frac{|t_b|^2 B}{|\Xi|^2},\\
&B= |t_b|^2 \cosh^2(\beta_0)(|z_s|^2+|g_s|^2+|f_s|^2+|\bar{f}_s|^2) \nonumber\\
&+2(2-|t_b|^2)[ \textrm{Re}(g_s z_s^\ast)\textrm{Re}(\cos\theta_0) +
\textrm{Re}(f_s z_s^\ast)\textrm{Re}(\sin\theta_0)] \nonumber\\
&+ 2|t_b|^2 \textrm{Im}(\cos\theta_0 \sin^\ast \theta_0) \textrm{Im}(f_s g_s^\ast).
\end{align}
The second term in Eq.~(\ref{ohm}) is the resistance of a normal 
conductor modified by penetrating Cooper pairs.
In what follows, we focus on the transport property at zero bias (i.e., $V_{\mathrm{bias}}=0$).
The solution of the Usadel equation in Eq.~(\ref{usadel}) at $\epsilon=0$ becomes
$\theta(x) = \theta_0(1+x/L)$ under the first boundary conditions in Eq.~(\ref{bc}).

The unconventional pair potentials in Eq.~(\ref{pair_symmetry}) have nodes on the Fermi surface. 
Such nodal superconductors cannot be straightforwardly classified into the ten well known topological classes~\cite{altland,schnyder} .
To characterize a nodal superconductor topologically, we consider the one-dimensional Brillouin zone
by fixing $k_y$ in the clean limit and define the one-dimensional winding number $w_{\mathrm{1D}}$~\cite{ms1dwn}. 
We find that $w_{\mathrm{1D}}=s_+$ for $s_+s_-=-1$ and $w_{\mathrm{1D}}=0$ for $s_+s_-=1$ with
$s_\pm= \Delta_\pm/|\Delta_\pm|$. Thus $w_{\mathrm{1D}}$ depends on the propagation channel.
The three unconventional pair potentials in Eq.~(\ref{pair_symmetry}) 
 satisfy $s_+s_-=-1$ for all the propagating channels. 
Therefore, such an unconventional superconductor hosts dispersionless zero-energy states (ZESs) at its 
clean surface~\cite{buchholtz,hara,hu,tanaka95}. 

The effects of such flat ZESs on the conductance depends on the parity of the pair potential. 
With spin-singlet even-parity superconductors, we can easily find 
that $I_F$ and $\theta(x)$ are real numbers. As a consequence, the second 
term in Eq.~(\ref{ohm}) becomes $R_{\mathrm{N}}$. Indeed, the solution is $\theta_0=0$ for a $d_{xy}$-wave.
In the limit of $R_{\mathrm{N}} \to \infty$, the solution for an $s$-wave is $\theta_0=\pi/2$. 
Therefore, the zero-bias differential conductance $G_{\mathrm{NS}}
= R_{\mathrm{NS}}^{-1}$ becomes
\begin{align}
\lim_{R_{\mathrm{N}} \to \infty} G_{\textrm{NS}} \to 0, \label{gns_singlet}
\end{align}
for all spin-singlet even-parity superconductors. 
By contrast, in the spin-triplet odd-parity superconductors,  
we find that
\begin{align}
I_F=& i \;{\sum_{k_y}} s_+ \frac{1-s_+s_-}{2} = i\, N_{\mathrm{ZES}}, \label{if_def}\\
N_{\mathrm{ZES}}=&N_+-N_-.\label{index}
\end{align}
The factor $(1-s_+s_-)/2$ extracts the propagating channels, each of which hosts 
a topologically protected ZES.
The integer $N_\pm$ corresponds to the number of ZESs characterized by $w_{\mathrm{1D}}=s_+ =\pm 1$.
In Eq.~(\ref{index}), the integer $N_{\mathrm{ZES}}$ is defined by the difference 
between $N_+$ and $N_-$. 
We obtain 
\begin{align}
\theta_0=& i \beta_0, \quad \beta_0=2 G_Q\, N_{\mathrm{ZES}} \, R_{\mathrm{N}}, \label{theta0}\\
I_B=&  J_1 \cosh^2(\beta_0) - N_{\mathrm{ZES}}\, \sinh(2\beta_0),\\
J_1 =& 2 \sum_{k_y}\left[ \frac{|t_b|^4}{(2-|t_b|^2)^2}\frac{1+s_+s_-}{2} +\frac{1-s_+s_-}{2}\right].
\end{align}
The resistance in Eq.~(\ref{ohm}) results in
\begin{align}
R_{\textrm{NS}} = \frac{1}{G_QI_B} + \frac{1}{2 G_Q N_{\mathrm{ZES}}} \tanh \beta_0.
\label{rns_triplet}
\end{align}
When we consider the limit of $R_{\mathrm{N}}\to \infty$, the first term vanishes and 
$\tanh \beta_0 \to \mathrm{sgn}(N_{\mathrm{ZES}})$. Thus we conclude that
\begin{align}
\lim_{R_{\mathrm{N}} \to \infty} G_{\textrm{NS}} \to \frac{4e^2}{h} \left|N_{\mathrm{ZES}}\right|,
\end{align}
for spin-triplet odd-parity superconductors. The minimum value of the zero-bias conductance 
is quantized. This is the main conclusion of this paper.

Next we consider the physical meaning of $N_{\mathrm{ZES}}$ by using the chiral symmetry of the
Hamiltonian. The BdG Hamiltonian in Eq.~(\ref{bdg}) satisfies
\begin{align}
 \left\{ \hat{H}_{\mathrm{BdG}}, \hat{\Lambda} \right\}_+ =0, \quad \hat{\Lambda}=\left[\begin{array}{cc}
 0 & i\\ -i & 0 \end{array}\right], \label{chiral}
\end{align}
which represents the chiral symmetry of the Hamiltonian. 
The eigenvalue of $\hat{\Lambda}$ is either $\lambda=1$ or $\lambda=-1$.
The eigenstates of $\hat{H}_{\mathrm{BdG}}$ have a characteristic property summarized as follows~\cite{ms1dwn}.

\noindent (i) A zero-energy state of $\hat{H}_{\mathrm{BdG}}$ is 
simultaneously an eigenstate of $\hat{\Lambda}$.
Namely, $\hat{\Lambda}\, \chi_\pm = \pm \, \chi_\pm$ holds for $\chi_\pm$ satisfying 
$\hat{H}_{\mathrm{BdG}}\, \chi_\pm=0$.  

\noindent (ii) On the other hand, nonzero-energy states are described by 
the linear combination of two states: one belongs to $\lambda=1$ and the other 
belongs to $\lambda=-1$.
Namely $ \chi_{E\neq 0} = a_+ \chi_{+} + a_- \chi_{-}$.
Moreover the relation $|a_+|=|a_-|$ always holds~\cite{si15}. 

\noindent In what follows, we show that $|N_{\mathrm{ZES}}|$ is the number of 
perfect transmission channels in a dirty normal metal while taking these properties into account.

 By deleting the normal segment $x<0$ 
 in Fig.~\ref{fig1}(a), we consider the surface of a clean semi-infinite superconductor. 
 The wave function of a ZES localized at the surface
 can be represented for each propagating channel,  
\begin{align}
\phi_{k_y} (\boldsymbol{r}) = A_{k_y} \left[ \begin{array}{c} i \\ s_+ \end{array} \right]
\sin(k_x x) e^{-x/\xi_\gamma} e^{ik_y y},
\end{align}
 where $A_{k_y}$ is the normalization constant and $\xi_\gamma=\hbar^2 k_F\cos\gamma/( m |\Delta_+|)$ 
depends on the pair potential in Eq.~(\ref{pair_symmetry}).
As suggested by property (i), $\phi_{k_y}$ is the eigenstate of $\Lambda$ 
belonging to its eigenvalue $\lambda = s_+$. 
We emphasize that the chiral eigenvalue $\lambda$, the sign of the pair potential 
$s_+$, and the one-dimensional winding number $w_{\mathrm{1D}}$ are identical to one another. 
Namely, the relation $w_{\mathrm{1D}}=s_+=\lambda$ holds for a propagation channel with $s_+s_-= -1$.
Therefore, in Eq.~(\ref{index}), $N_\pm$ is exactly equal to the number of ZESs that belong to 
$\lambda=\pm 1$. Mathematically, $N_{\mathrm{ZES}}$ corresponds to the Atiyah-Singer index which 
is an invariant as far as the Hamiltonian preserves the chiral symmetry in Eq.~(\ref{chiral}).
At a clean surface, the degree of degeneracy 
at zero energy is $N_c$ for three unconventional pair potentials in Eq.~(\ref{pair_symmetry}). 
The translational symmetry in the $y$ direction protects such a high degeneracy at zero energy.
When we attach a dirty normal metal to form an NS junction, however, 
the potential disorder lifts the degeneracy at zero energy. Even in a dirty NS junction, 
$N_{\mathrm{ZES}}$ remains unchanged because the impurity potential preserves the chiral symmetry.
We first count $N_{\mathrm{ZES}}$ at a clean surface of the superconductor. 
Then we discuss how the potential disorder lifts the degeneracy depending on $N_{\mathrm{ZES}}$.

With a $p_{x}$-wave, we find $N_+= N_c=N_{\mathrm{ZES}}$ and $N_-=0$ because
 $\lambda=1$ for all the propagation channels.
Such a pure chiral state cannot form nonzero-energy states according to property (ii) 
because ZESs belonging to $\lambda=-1$ are absent.
The degree of degeneracy remains unchanged from 
$N_c=N_{\mathrm{ZES}}$ in the dirty normal metal~\cite{si15}. 
 This fact explains the anomalous proximity effect~\cite{yt04,ya06,ya07}. 
With a $d_{xy}$-wave, $N_+= N_-$ is satisfied because the ZESs for 
$k_y >0$ ($k_y <0$) belong to $\lambda=1$ ($\lambda=-1$). 
Thus we find $N_{\mathrm{ZES}}=0$.
The impurity potential completely eliminates the degeneracy at zero energy, which describes 
the absence of the proximity effect in a $d_{xy}$-wave NS junction~\cite{ya01,yt03}.
The conclusion $N_{\mathrm{ZES}}=0$ is valid for all spin-singlet even-parity superconductors.
Actually $I_F$ in Eq.~(\ref{bc}) is real for all spin-singlet superconductors, 
whereas $N_{\mathrm{ZES}}$ is defined by the imaginary part of $I_F$ in Eq.~(\ref{if_def}).
Finally, with a $f$-wave, we obtain $N_+=[ (Wk_F/\pi)/ \sqrt{2} ]_{\mathrm{G}}$ and 
$N_-=[ (Wk_F/\pi)(1- 1/\sqrt{2})]_{\mathrm{G}}$ by using Eq.~(\ref{kysum}). 
The number of ZESs in the dirty normal metal is $N_{\mathrm{ZES}}=[ (Wk_F/\pi)(\sqrt{2}-1)]_{\mathrm{G}}$ 
because ZESs with $\lambda=1$ and $\lambda=-1$ couple one-by-one
and form two nonzero-energy states according to property (ii).

 The integer $|N_{\mathrm{ZES}}|$ in Eq.~(\ref{index}) 
indicates the number of ZESs remaining in a dirty normal metal. 
Under the boundary condition in Eq.~(\ref{bc}), 
the imaginary part of $\theta(x)$ is proportional to $N_{\mathrm{ZES}}$.
The local density of states (LDOS) is given by $\rho(x,\epsilon) = \rho_0 \mathrm{Re}[\cos\theta(x,\epsilon)]$ 
with $\rho_0$ being the density of states in the normal state at the Fermi energy. 
The resulting LDOS at $\epsilon=0$ 
\begin{align}
\rho(x)/\rho_0 = \cosh[ 2 G_Q\, N_{\mathrm{ZES}} \, R_{\mathrm{N}} \, (1+x/L)] \gg 1,
\end{align}
is greatly enhanced, which
reflects the penetration of the ZESs into a normal metal.
The penetrating ZESs form resonant transmission channels whose number is $|N_{\mathrm{ZES}}|$. 
As a consequence, $|N_{\mathrm{ZES}}|$ gives the quantized minimum value of the conductance in the strong scattering limit.

We check the validity of the above argument 
by employing a numerical simulation on the two-dimensional single-band tight-binding model.
We choose $L=30a_0$, $W=25a_0$, $\mu_F/t=1.0$, $\Delta/t=0.01$ 
with $a_0$ and $t$ being the lattice constant and the hopping integral, respectively.
This parameter choice leads to $N_c=9$.
In the presence of the random potential, we plot the $G_{\mathrm{NS}}$ versus 
$G_{\mathrm{N}}=R_{\mathrm{N}}^{-1}$ in Fig.~\ref{fig1}(b), where the two conductances are calculated  
independently by using the recursive Green function method~\cite{lee,ando}.
As predicted by the quasiclassical Green function method, $G_{\mathrm{NS}}$ is 
quantized at $2 G_Q N_{\mathrm{ZES}}$ for spin-triplet junctions.
For a $p_x$-wave symmetry, $N_+=N_c=9$ and $N_-=0$ are obtained numerically, which results in
$N_{\mathrm{ZES}}=9$. For an $f$-wave symmetry, we find $N_{\mathrm{ZES}}=5$ because $N_+=7$ and $N_-=2$.
The results also show that $G_{\mathrm{NS}}$ goes to zero with a decrease in 
$G_{\mathrm{N}}$ for spin-singlet $s$- and $d_{xy}$-wave cases. 
The numerical results justify our conclusions.

The nonzero conductance minimum is a character of 
odd-parity superconducting states that have been proposed in an artificial thin film~\cite{alicea} 
and exotic materials such as (TMTSF)$_2$X (X = PF$_6$, ClO$_4$, etc.)~\cite{jerome,bechgaad,kuroki} and 
Na$_x$CoO$_2\cdot$ yH$_2$O~\cite{takeda,kuroki2}. 
Thus the contents of this paper have a strong connection to the physics of Majorana fermion~\cite{majorana}.
Actually, the conductance quantization~\cite{lutchyn,oreg,mourik,deng,das,ya13} in a nanowire superconductor
is an example of the present theory. Moreover
our analysis using the chiral symmetry of the Hamiltonian is closely related to
an argument on the stability of Majorana flat bands~\cite{querioz}.

In summary, we have discussed the zero-bias 
differential conductance $G_{\mathrm{NS}}$ in a normal-metal/superconductor junction consisting of
a nodal superconductor preserving time-reversal symmetry. 
The minimum value of $G_{\mathrm{NS}}$ is quantized at $(4e^2/h)N_{\mathrm{ZES}}$.
The analysis in terms of the chiral symmetry of the Hamiltonian indicates that 
the integer $N_{\mathrm{ZES}}$ is the Atiyah-Singer index in mathematics.
%One remaining issue of this paper may be to make clear the relation between  
%the Atiyah-Singer's index and a topological invariant defined in terms of the scattering 
%matrix~\cite{dahlhaus}.

%\begin{acknowledgments}
The authors are grateful to M. Sato, Ya. V. Fominov, and A. A. Golubov for useful discussions.
This work was supported by ``Topological Materials Science'' (Nos.~15H05852 and 15H05853) 
and KAKENHI (Nos.~26287069 and 15H03525) from
the Ministry of Education,
Culture, Sports, Science and Technology (MEXT) of Japan
and by the Ministry of Education and Science of the Russian Federation
(Grant No.~14Y.26.31.0007).
%\end{acknowledgments}


\end{document}